\documentclass[twocolumn,aps,prl,showpacs,superscriptaddress]{revtex4}

\usepackage{color}
\usepackage{mathtools}
\usepackage{graphicx}
\usepackage{SIunits}

\begin{document}

\title{{Unconventional Josephson junctions with topological Kondo insulator weak links}}

\author{Xuecheng~Ye}
\affiliation{Department of Physics and Materials Research Center, Missouri University of Science and Technology, Rolla, MO 65409}

\author{Jacob~Cook}
\affiliation{Department of Physics and Materials Research Center, Missouri University of Science and Technology, Rolla, MO 65409}

\author{Erik~D.~Huemiller}
\affiliation{Department of Physics and Materials Research Laboratory, University of Illinois at Urbana-Champaign,
Urbana, Illinois 61801}

\author{Aaron~D.~K.~Finck}
\affiliation{IBM T.~J. Watson Research Center, Yorktown Heights, NY 10598}

\author{Pouyan Ghaemi}
\affiliation{Physics Department, City College of the City University of New York, New York, NY 10031}

\author{Thomas~Vojta}
\affiliation{Department of Physics and Materials Research Center, Missouri University of Science and Technology, Rolla, MO 65409}

\author{Vivekananda Adiga}
\affiliation{IBM T.~J. Watson Research Center, Yorktown Heights, NY 10598}

\author{Shanta~R.~Saha}
\affiliation{Department of Physics, University of Maryland, College Park, MD}

\author{Johnpierre~Paglione}
\affiliation{Department of Physics, University of Maryland, College Park, MD}

\author{Cihan~Kurter}
\thanks{Current Affiliation:~IBM T.~J. Watson Research Center, Yorktown Heights, NY 10598}
\email{cihan.kurter@ibm.com}
\affiliation{Department of Physics and Materials Research Center, Missouri University of Science and Technology, Rolla, MO 65409}

\date{\today}

\begin{abstract}

Proximity-induced superconductivity in three dimensional (3D) topological insulators forms a new quantum phase of matter and accommodates exotic quasiparticles such as Majorana bound states. One of the biggest drawbacks of the commonly studied 3D topological insulators is the presence of conducting bulk that obscures both surface states and low energy bound states. Introducing superconductivity in topological Kondo insulators such as SmB$_6$, however, is promising due to their true insulating bulk at low temperatures. In this work, we develop an unconventional Josephson junction by coupling superconducting Nb leads to the surface states of a SmB$_6$ crystal. We observe a robust critical current at low temperatures that responds to the application of an out-of-plane magnetic field with significant deviations from usual Fraunhofer patterns. The appearance of Shaphiro steps under microwave radiation gives further evidence of a Josephson effect. Moreover, we explore the effects of Kondo breakdown in our devices, such as ferromagnetism at the surface and anomalous temperature dependence of supercurrent. Particularly, the magnetic diffraction patterns show an anomalous hysteresis with the field sweep direction suggesting the coexistence of magnetism with superconductivity at the SmB$_6$ surface. The experimental work will advance the current understanding of topologically nontrivial superconductors and emergent states associated with such unconventional superconducting phases.

\end{abstract}


\maketitle


Topological superconductivity~\cite{SatoReview} is expected to be a unique platform to generate and manipulate zero energy modes, referred to as Majorana bound states~\cite{Majorana1937,NatPhys.5.614,Alicea2012,Beenakker2013, Grosfeld19072011,SST.27.124003}. One way to get a topological superconductor is engineering Josephson junctions on the surfaces of 3D topological insulators~\cite{PhysRevLett.100.096407, PhysRevLett.109.056803, KurterNature, Orlyanchik2013, NatCommun.4.1689, PhysRevB.85.045415, NatMat.11.417, SciRep.2.339,NatCommun.2.575,PhysRevB.84.165120}. In such devices, the quantum interference of electron and hole-like excitations will form low energy Andreev bound states  whose spectrum is sensitive to the relative phase difference between the superconducting leads and the microscopic details of the barrier. Proximity-induced supercurrent flowing through the topological insulator segment of the junction is carried by such states.

Initially, Josephson junctions incorporating Bi-based 3D topological insulators were favored for experimental searches for signatures of Majorana modes. However, significant bulk and trivial surface state contribution to the electronic transport complicated the interpretation of such experiments. Although electrostatic gating can alleviate the problem~\cite{Steinberg, NatCommun.2.575,KurterPRB}, the quest for finding a proper platform for such modes is still ongoing.  

Topological Kondo insulators~\cite{PhysRevLett.104.106408, PhysRevB.85.045130} such as SmB$_6$ are promising candidates for solving the problems caused by trivial transport channels. At high temperatures, these materials are metallic with a dense array of magnetic moments from $f$-electrons. However, at sufficiently low temperatures, such moments strongly couple to conduction electrons leading to the formation of singlets. This hybridization opens up a narrow gap in the electronic band structure~\cite{Dzero_Rev}. Point contact tunneling measurements confirmed the formation of such a gap below 60 K~\cite{PhysRevX.3.011011}.  At even lower temperatures (below 3-5 K), an anomalous saturation of sample resistance was observed~\cite{PhysRevB.20.4807}, which has been attributed to the presence of topological surface states ~\cite{Fisk1, Fisk2, Paglione}.  Recent low temperature transport studies with SmB$_6$ have revealed various anomalies such as thickness independence of Hall voltage, suggesting that transport is dominated by surface states~\cite{SciRep.3.3150}. There has been an evidence of helical nature of such states based on electrical detection of surface spin polarization~\cite{arXiv:1809.04977}. Moreover, the observation of a perfect Andreev reflection in a Au-SmB$_6$/YB$_6$ structure has been claimed due to the topologically protected surface states as well as the absence of a bulk conduction channel~\cite{Perfect}. In addition, photoemission~\cite{NatCommun.5.4566} and STM~\cite{PhysRevLett.112.136401} studies supported the presence of in-gap surface states that are topologically protected.



We studied a set of single lateral Josephson junctions on the surface of mechanically polished SmB$_6$ crystals. In this paper we will show results from four different junctions labelled as junction-1 through junction-4. The junction lengths vary from 50 nm to 200 nm. The cubic crystal structure of SmB$_6$ does not permit exfoliation to obtain atomically thin flakes.  Here, the entire circuit of the devices (junctions  and contact leads/pads) is fabricated on the surface of the mechanically polished single crystals.

\begin{figure}
\centering
\includegraphics[bb=1 1 721 324,width= 3.5 in]{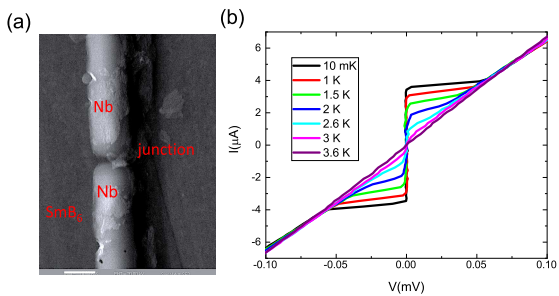}
\caption{(Color online) (a)~Scanning electron microscopy SEM image of one of the studied devices, junction-1, made with two Nb leads $\sim$ 100 nm apart on a polished crystal of SmB$_6$. (b) Current-voltage I-V characteristics of the junction-1 for a set of temperatures. The supercurrent persists up to 3.6 K. } \label{fig:SEM}
\end{figure}

An SEM picture of one of the studied devices, junction-1 is shown in Fig.~\ref{fig:SEM}(a). Two superconducting Nb leads that are about 100 nm apart have been defined by electron beam lithography and $\sim$  60-70 nm Nb deposition via magnetron sputtering. Prior to Nb deposition an in-situ Ar milling was applied to lightly etch the top surface of the crystal and to ensure good interface between SmB$_6$ and the superconductor.

All of our devices were thermally anchored to the mixing chamber of a cryogen-free dilution refrigerator with a base temperature of 10 mK and equipped with filtered wiring. The transport measurements were done with standard lock-in techniques at different temperatures and magnetic fields. Figure~\ref{fig:SEM}(b) shows the temperature dependence of current-voltage (IV) characteristics of the junction-1, demonstrating a clear induced supercurrent at 10 mK. Similar to other samples, there is no significant change in the supercurrent up to $\sim$ 300-400 mK. This suggests that Kondo hybridization is strong in this temperature regime where we believe that the singlets formed on the surface states predominantly carry the supercurrent. Beyond that regime, we see a monotonic drop in the critical current (I$_c$) until all signs of induced superconductivity vanishes. For most of the devices, supercurrent survives up to 3-6 K~\cite{PhysRevX.6.031031}. This differs in proximity-induced Nb/Bi$_2$Se$_3$ devices where superconductivity often is suppressed well below the T$_c$ of the Niobium, approximately beyond 1 K~\cite{NatCommun.4.1689, Orlyanchik2013}. 


\begin{figure}
\centering
\includegraphics[bb=1 1 911 830,width= 3.5 in]{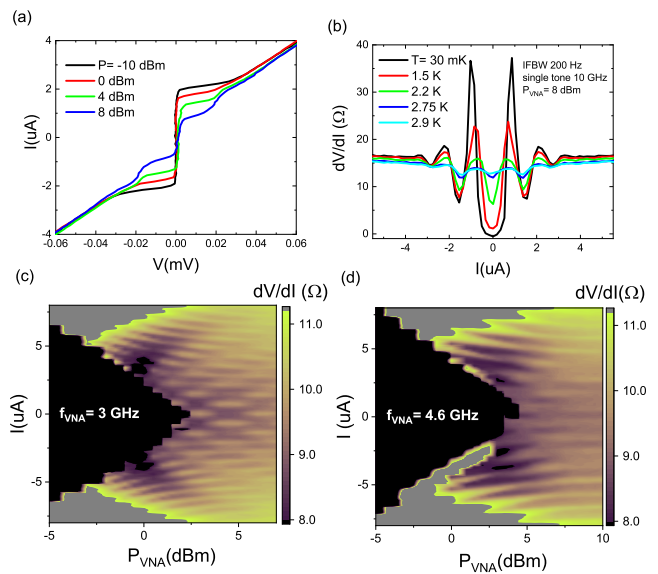}
\caption{(Color online) (a) AC Josephson effect giving rise to Shaphiro steps in IV  and (b) corresponding peaks in differential resistance for another sample, junction-1. (c) and (d) Color plots demonstrating such steps for junction-2 at two different frequencies.} \label{fig:ACJos}
\end{figure}

To confirm the presence of the Josephson effect in our junctions, we performed AC Josephson effect measurements by means of microwave irradiation. For a junction with 2$\pi$ periodic Josephson relation, microwaves applied at a frequency of $f$ gives rise to stair-like features in the IV characteristics with voltage spacing of $\Delta V = hf/2e$. These features are known as Shaphiro steps and correspond to minima in differential resistance vs bias current measurements~\cite{PhysRevLett.121.097701}. 

To observe Shaphiro steps, the junctions are irradiated by microwaves by means of a coaxial cable whose center pin is about 1 mm vertically away from the surface of the sample. Figure~\ref{fig:ACJos}(a) shows IV characteristics for a set of microwave power values for junction-1. As the microwave power is increased, the Shaphiro steps start to appear beyond 4 dBm that survive up to 3 K as seen from the temperature dependence measurements of dV/dI of the same junction in Fig.~\ref{fig:ACJos}(b). The voltage spacing of the Shaphiro steps is 20 $\mu$V as expected from the applied 10 GHz microwave tone, signifying that the current-phase relationship is $2\pi$ periodic.

We observed Shaphiro steps in multiple devices; Figs.~\ref{fig:ACJos}(c) and (d) show color plots of dV/dI vs bias current and microwave power for junction-2 at two different frequencies, 3 and 4.6 GHz respectively. Upon increasing microwave power, the $I_c$ monotonously decreases and finally vanishes at roughly 2.5 dBm for 3 GHz and 5 dBm for 4.6 GHz.  Beyond these power levels $I_c$ starts to oscillate with higher power supporting the 2$\pi$ periodic AC Josephson effect further.

\begin{figure}
\centering
\includegraphics[bb=1 1 860 715,width= 3.5 in]{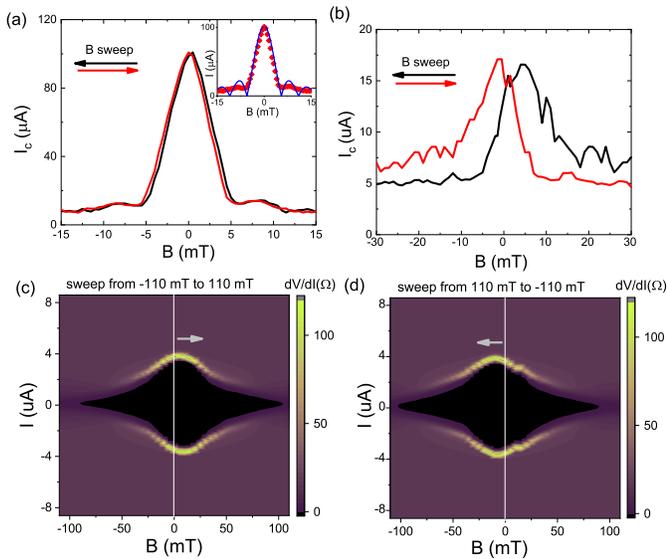}
\caption{(Color online) (a), (b) Magnetic diffraction patterns of the I$_c$ for the junction-3 and the junction-2 as the out of plane magnetic field is swept in forward and reverse directions. (c) and (d) Color plots of dV/dI vs bias current as a function of perpendicular magnetic field for the junction-1, clearly showing a discernible shift of I$_c$ with respect to the sweep direction.} \label{fig:mag_field}
\end{figure}

Now we turn to the out-of plane magnetic field response of the supercurrent. Figure~\ref{fig:mag_field}(a) shows magnetic field oscillations of the I$_c$ for the junction-3 at 10 mK that is similar to a Fraunhofer pattern. This pattern is another characteristic of the Josephson effect and is generated by quantum interference from phase winding induced by the magnetic flux within the junction.  When a perpendicular field is applied to a conventional junction the phase will vary along the barrier, thus the supercurrent  through  the  barrier  will modulate with flux according  to  I$_c(\Phi)$=  I$_c$(0) $\mid$(sin($\pi \Phi$)/$\Phi_0$)/($\pi \Phi$)/$\Phi_0$)$\mid$ where $\Phi_0$ is magnetic flux quantum. The inset of the Fig.~~\ref{fig:mag_field}(a) shows a theoretical Fraunhofer pattern plotted with the measured data acquired with increasing magnetic field sweep, demonstrating a reasonable agreement between the two.

The magnetic field response of I$_c$ is quite different for junction-2, as shown Fig~\ref{fig:mag_field}(b).  First, there is a distinct lack of side lobes in the diffraction pattern demonstrating significant deviations from a conventional Fraunhofer pattern. Second, the maximum critical current occurs at about $\pm$ 5 mT instead of zero field. We attribute this shift in applied field axis with respect to the ideal Fraunhofer diffraction pattern to the existence of magnetization in the SmB$_6$ surface states, which generates additional flux that must be cancelled out by an applied magnetic field in order to observe maximal critical current. The origin of such magnetism will be discussed later in the text. Although the direction where the central peak of diffraction pattern shifts is unexpected, similar hysteresis was observed in Sr$_2$RuO$_4$ due to multiple, dynamical domains of order parameter generating chiral supercurrents~\cite{DaleScience}.  

As we were performing these hysteresis measurements,  we observed discernible suppressed supercurrent at zero field when the magnetic field is ramped down from a positive value. To check whether such suppression might be due to flux trapping or vortex entry, the fridge was warmed up to 20 K, well above the critical temperature of Nb and the superconducting magnet and then cooled back down to base temperature again. Before applying any magnetic field, we observed similar suppressed critical current at zero field which confirmed that the hysteresis in our diffraction patterns is not due to trapped vortices but possibly due to ferromagnetic behavior of surface states in SmB$_6$.  The maximum critical current was only revived after sweeping the magnet in the opposite polarity.

The color plots of dV/dI vs bias current and applied magnetic field for junction-1 at 10 mK are shown in Fig.~\ref{fig:mag_field}(c) and (d). Once again the diffraction pattern lacks side lobes.  However, here the shift is in the opposite direction as for junction-2. Previously, superconductor-ferromagnet-superconductor (SFS) Josephson junctions showed anomalous Fraunhofer patterns with hysteretic behavior similar to data shown in Fig.~\ref{fig:mag_field}(c) and (d); demonstrating maximal supercurrent occurring at nonzero applied field~\cite{PhysRevB.79.094523}.  More recently, hysteretic magnetotransport has been observed in topological systems such as magnetically doped Bi$_2$Se$_3$ ~\cite{checkelsky} and SmB$_6$, which have been also associated with ferromagnetic domain walls in surface states~\cite{Paglione}.

\begin{figure*}
\centering
\includegraphics[bb=4 6 1052 343,width= 5.5 in]{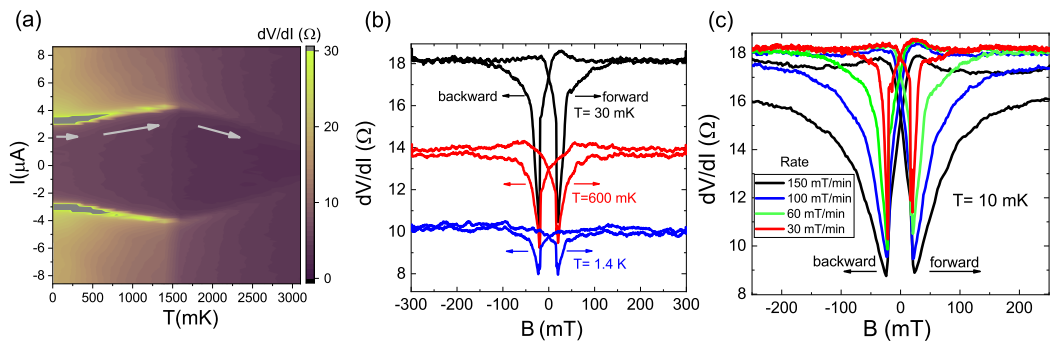}
\caption{(Color online) (a) Color plot of dV/dI vs bias current as a function of temperature for junction-4. (b) Temperature evolution of hysteresis in normal state resistance with sweep direction for junction-4. (c) Rate dependence of magnetic field hysteresis in normal state resistance of junction-4.  } \label{fig:anamoly}
\end{figure*}

At low temperatures, we expect the formation of Kondo singlets in SmB$_6$ without magnetic behavior.  We can explain the appearance of ferromagnetic behavior by invoking a Kondo breakdown~\cite{PhysRevLett.114.177202, PhysRevLett.116.046403}, which liberates a large number of randomly oriented magnetic moments that were previously coupled to conduction electrons inside Kondo singlets. When we have a significant population of $f$-electrons, the material can have magnetic behavior even at mK temperatures.  The breakdown of the Kondo effect at the layers close to the boundary of the sample stems from the reduced screening of the local moments due to broken translational symmetry at the outmost layers. This leads to a major modification in the band structure and reduces the Kondo temperature significantly.

When the Kondo hybridization is strong and the surface states are not subject to Kondo breakdown, we obtain more conventional magnetic diffraction patterns with no sensitivity to field sweep direction as shown in Fig.~\ref{fig:mag_field}(a). As the spins are freed from the Kondo singlets due to the breakdown, they can form ferromagnetic domains on the surface generating negative or positive flux within the junction barrier. This will lead to modifications in Fraunhofer patterns with the shifts of the maximum supercurrent towards either positive or negative magnetic field. Similar phenomenon was observed in Ref~\cite{DaleScience}.

Finally, in junction-4 we observe an anomalous temperature dependence of the I$_c$ that can also be explained by a Kondo breakdown. Figure~\ref{fig:anamoly}(a) shows a color plot of dV/dI vs temperature and bias current demonstrating the full evolution of I$_c$ as the sample is heated. Critical current shows distinctive behavior in three different temperature regimes. For very low temperatures (i.e. below 300 mK), it exhibits almost no change.  In this regime, we expect that Kondo hybridization is strong and that the supercurrent is carried purely by the protected surface states. Then the I$_c$ gradually increases between 300 mK and 1.6 K. Although this observation is unexpected from a usual Josephson junction, it could be consistent with the thermal population of trivial carriers either in the bulk or surface at higher temperatures that can carry additional supercurrent in parallel with the topological surface states (Kondo singlets) in SmB$_6$ triggered by the Kondo breakdown. Indeed, we observe the normal state resistance monotonically decreases with increasing temperature, as trivial states are being thermally activated. Furthermore, it is reported that especially the (001) surface of SmB$_6$ is polar~\cite{PhysRevLett.111.216402} which gives rise to various modifications of the surface states, such as band bending, formation of 2D electron gases and quantum well confinements~\cite{NatCommun.7.13762}. It is conceivable that different conditions at the surface could be caused by uncontrolled variations of the nanofabrication or sample/crystal preparation such as polishing. Beyond 1.6 K the critical current monotonously declines until the induced superconductivity is destroyed beyond 3-4 K. In this regime, the free magnetic moments due to Kondo breakdown are possibly too dominant to be screened by the Kondo singlets. Thermal dephasing of such moments can explain the rapid decrease in supercurrent. 

In the same junction we also observe unusual properties of the normal state. Normal state conductance has been studied by applying a bias current that is much larger than the critical current. The longitudinal magnetoresistance data at low temperatures and low fields exhibits a sharp suppression in normal state resistance (R$_N$) near zero field as seen in Fig.~\ref{fig:anamoly}(b). Intriguingly, when we change the sweep direction of the magnetic field, we observe a butterfly-shaped hysteretic behavior of magnetoresistance~\cite{Anomaly} with two separate minima at B$_{min}$ = $\pm$ 24 mT. This feature was previously attributed elsewhere to edge channels between ferromagnetic domains in SmB$_6$~\cite{Paglione}. However, one must also consider trivial explanations for such magnetic hysteresis, including magnetocaloric effects or magnetic impurity scattering on the surface of the material~\cite{PhysRevB.92.115110}. As we increase the temperature, the hysteretic signal gets weaker as the ferromagnetism is suppressed by thermal fluctuations.

The dip feature is reminiscent of data that appears at zero field magnetoresistance in spin-orbit coupled materials due to weak antilocalization effect, which is a correction to classical magnetoresistance arising from quantum interference of scatted electron waves. The clear hysteresis of the dips and magnetic sweep rate dependent R$_N$ as seen in Fig.~\ref{fig:anamoly}(c) suggest that the features emerging in low temperature magnetoresistance are not due to a weak antilocalization effect.

In conclusion, we created and studied an unconventional Josephson junction using the surface states of a Kondo insulator as a weak link between superconducting leads. The observed critical temperature of the induced supercurrent is much higher compared to other proximity-induced topological Josephson devices~\cite{PhysRevX.6.031031,PhysRevB.96.165408}. With the microwave irradiation of the junctions, we obtained clear Shaphiro steps in IV characteristics and dV/dI oscillations beyond the suppression of the supercurrent which are manifestations of the AC Josephson effect.  The junctions demonstrated hysteretic response to out-of-plane magnetic field. The ferromagnetism at the surface at low temperatures can be attributed to a Kondo breakdown which generates free magnetic spins by breaking the Kondo singlets. Similar hysteresis was also observed in the normal state magnetoresistance supporting the claims about magnetic properties of the outmost layers further.

$\it{Acknowledgements.}$ CK acknowledges funding by UMRB. For the device fabrication, we used the facilities of Missouri S$\&$T Materials Research Center. We would like to thank
Corey Buris and Clarisa Wisner for their assistance in device preparation. We thank David Pomerenke for microwave amplifiers, and Dale Harlingen for his ion beam facilities. We appreciate the discussions with Onur Erten and Joseph Glick.

\bibliography{SmB6}

\end{document}